\documentclass[10pt, conference]{IEEEtran}

\IEEEoverridecommandlockouts

\usepackage{enumitem}
\usepackage[colorinlistoftodos]{todonotes}
\usepackage{cite}
\usepackage{amsmath,amssymb,amsfonts}
\usepackage{algorithmicx}
\usepackage{array}
\usepackage{balance}
\ifCLASSOPTIONcompsoc
  \usepackage[caption=false,font=normalsize,labelfont=sf,textfont=sf]{subfig}
\else
  \usepackage[caption=false,font=footnotesize]{subfig}
\fi
\usepackage{url}
\usepackage{graphicx}
\usepackage{amsthm}
\usepackage{svg}
\usepackage{bm}
\usepackage{algorithm}
\usepackage[noend]{algpseudocode}
\usepackage{setspace}
\usepackage{xcolor,soul}
\usepackage{soul}
\usepackage{comment}

\usepackage[scaled=.85]{berasans}

\algdef{SE}[DOWHILE]{Do}{doWhile}{\algorithmicdo}[1]{\algorithmicwhile\ #1}%

\algrenewcommand\algorithmicindent{1em}%

\begin{document}

\title {Near Optimal VNF Placement in Edge-Enabled 6G Networks}
\author{
\IEEEauthorblockN{Carlos Ruiz De Mendoza}
\IEEEauthorblockA{
Universitat Politècnica de Catalunya (UPC) \\
carlos.ruiz.de.mendoza@estudiantat.upc.edu
}
\and

\IEEEauthorblockN{
    Bahador Bakhshi,
	Engin Zeydan, 
	Josep Mangues-Bafalluy
}
\IEEEauthorblockA{
Centre Tecnol\`{o}gic de Telecomunicacions de Catalunya (CTTC/CERCA)\\
\{bbakhshi, engin.zeydan, josep.mangues\}@cttc.cat
}

\thanks{{This  work  has  been  partially  funded  by  the EC  H2020  5Growth  Project  (grant  no.  856709). The work by Carlos Ruiz de Mendoza was done while carrying out the Master Thesis at CTTC.}}

}

\maketitle

\begin{abstract}
Softwarization and virtualization are key concepts for emerging industries that require ultra-low latency. This is only possible if computing resources, traditionally centralized at the core of communication networks, are moved closer to the user, to the network edge. However, the realization of Edge Computing (EC)  in the sixth generation (6G) of mobile networks requires efficient resource allocation mechanisms for the placement of the Virtual Network Functions (VNFs). Machine learning (ML) methods, and more specifically, Reinforcement Learning (RL), are a promising approach to solve this problem. The main contributions of this work are twofold: first, we obtain the theoretical performance bound for VNF placement in EC-enabled 6G networks by  formulating the problem mathematically as a finite Markov Decision Process (MDP) and solving it using a \textit{dynamic programming} method called \textit{Policy Iteration} (PI). Second, we develop a practical solution to the problem using RL, where the problem is treated with \textit{Q-Learning} that considers both computational and communication resources when placing VNFs in the network. The simulation results under different settings of the system parameters show that the performance of  the \textit{Q-Learning} approach is close to the optimal \textit{PI} algorithm (without having its restrictive assumptions on service statistics). This is particularly interesting when the EC resources are scarce and efficient management of these resources is required.\\
\end{abstract}

\begin{IEEEkeywords}
NFV, EC, Dynamic Programming, MDP,  \textit{Policy Iteration}, \textit{Q-Learning}, Reinforcement Learning 
\end{IEEEkeywords}

\section{Introduction}

Future beyond 5G(B5G)/6G networks are expected to bring new improvements over the previous mobile networks (4G and 5G) and enable new paradigms. This technological evolution paves the way for endless services tailored to specific verticals; e.g., e-Health, Industry 4.0, Internet of Things (IoT), and automotive. More cells and antennas will be deployed in combination with advanced technologies, such as Virtual Radio Access Network (vRAN), which enable partial or full virtualization of the network through Network Function Virtualization (NFV) and Network Slicing. To serve all these needs, current and future networks heavily rely on the Edge Computing (EC) concept to reduce the distance between the end user and the computing resources. By placing Virtual Network Functions (VNF) in proper edge computing centers at the edge of the network, data is no longer stored or processed in a distant data center, which translates into a significantly lower latency \cite{corneo2021much} at the cost of additional complexity of network management.

B5G/6G technologies are expected to provide an increase in capabilities, all in a multi-vendor environment, with particular focus on ultra-low latency data transmission. It is no wonder that B5G/6G networks are expected to become complex systems, making it difficult to deploy and manage services on them. The options abundance in  terms  of offered services makes it even more difficult to determine an optimal way to deliver them in the  future and to predict the demand in order to deploy in the most efficient way the infrastructure. Such a complex system requires advanced methods targeting an optimal use of its resources.

A major issue with managing EC-enabled 6G network is the inter-relation between the computing and communication resources to provide services---to admit the user's request, the corresponding VNF should be placed in a EC center that has sufficient computing resources as well as sufficient communication resource to handle the offered traffic load. Therefore, dealing with the edge cloud and transport Key Performance Indicators (KPIs) simultaneously is the added complexity for the MNO when purporting to provide scalability. This is one of the issues that we studied in this paper.


Furthermore, in this paper we also analytically find the optimal performance bounds of such edge cloud scenarios that, unlike previous work, combine computing and communication characteristics to take decisions. The scenario is designed so that it is mathematically tractable. Consequently, the problem is formulated as a Markov Decision Process (MDP) and its optimal solution is found through the \textit{Policy Iteration} (PI) algorithm. After that, this paper focuses on the design of a practical algorithm that is able to perform as close as possible to the theoretical bound in terms of rejection ratio. It is practical in the sense that it does not take any restrictive assumption on service statistics, hence it can be applied to real networks. And given its low rejection ratio compared to other approaches, it makes the most out of the resources of the operator. In this direction, and given their potential, AI/ML approaches \cite{5GPPPaper}, and more specifically, reinforcement learning, is shown to perform close to optimal under multiple conditions if enough iterations are run.

\subsection{Related Work}

Resource management between network edge nodes has been widely researched and studied in various domains. The authors in \cite{8885745} propose a Q-Learning algorithm to choose optimal offloading among Fog network nodes. To evaluate the performance of this algorithm, it is compared with existing offloading methods as: Least-Queue, Nearest Node or Random Node selection. Reference \cite{mecoff} formulates the resource allocation problem in EC as a minority game, and then compares the performance of different RL methods to make its agent solve the game. Akkarit et al. \cite{inproceedings} present the automatic adaptation of container instances under a Q-Learning algorithm and also with the implementation of neural networks to maintain a certain service quality level without reducing the cloud computing resources. Yala L. et al. \cite{8647858} propose an algorithm for placing VNF  using  optimization-adapted Genetic Algorithm meta-heuristic which aims at minimizing latency and maximizing service availability. 

Some of the above works differ from this paper in that the evaluation stage of the proposed algorithms are compared with solutions that do not necessarily yield the optimal solution which considers both transport and cloud related parameters. In this paper, \textit{Q-Learning} is evaluated against a mathematical solution that allows measuring the performance gap between the two. Moreover, the problem is posed as a finite MDP and solved by a model-based and a model-free RL algorithm.

\vspace{-1mm}
\subsection{Main Contributions}
\label{sec_related}

This paper takes the architecture of the H2020 5GPPP 5Growth project\footnote{https://5growth.eu/} \cite{li20215growth} as a general reference. The 5Growth project aims to validate the operation of 5G systems deployed in vertical industries and incorporates (through open APIs) AI/ML-related algorithms in service deployment and operation. 

In this paper, VNF placement decisions for each service requests are made according to specific operational requirements (e.g., ensuring an efficient use of edge resources while maximizing the number of services delivered over the shared infrastructure).
First, the theoretical performance bounds are calculated through Dynamic Programming (DP) and in the form of a model-based MDP solved through the PI algorithm. For this, all state transition probabilities must be computed in advance, as well as all possible valid states. This is an algorithm that mathematically allows the optimal solution to be obtained from all possible solutions in the search space.
Another algorithm representing reasonable operation guidelines is \textit{best fit}, which is also taken as reference for comparison with theoretical bounds and practical learning algorithms. This algorithm is inspired in the classic load balancing algorithm Weighted Round Robin \cite{articlewrr}, but adapted to the needs of the topic under consideration. In \textit{best fit} approach, each EC has a weight based on the administrators chosen criteria. The EC with the highest weight serves the request.

Finally, the problem has been approached through a \textit{Q-Learning off-policy time differential} algorithm to conceive a deployable algorithm in practice that performed as close as possible to the optimal bound. In this case, an agent observes each incoming new VNF request. Based on the network state, the agent performs an action by assigning it to an optimal EC node to maximize the total number of processed VNF requests in the system. 
In summary, the key contributions of this paper are:

\begin{itemize}
    \item The optimal VNF placement problem is formulated as a finite MDP and solved using a model-based algorithm, i.e., PI, considering cloud and transport network conditions, hence obtaining the optimal performance bounds.
    \item A practical solution is given for a (near) optimal solution using a model-free and off-policy algorithm, i.e., \textit{Q-Learning}, and compared with PI and \textit{best fit}.
    \item The simulation results make it clear that \textit{Q-Learning} works near-optimally when EC resources must be managed conscientiously.
\end{itemize}

The rest of the paper is organized as follows. Section \ref{scenario} describes the considered scenario, system model and the problem statement. Section \ref{sec_MDP} presents the problem as an MDP and the PI approach. Section \ref{sec_QLRL} introduces both practical solutions, \textit{Q-Learning} and \textit{best fit} algorithms. Section \ref{sec_sim} details each of the simulations that have been carried out. Finally, section VI draws the conclusions.


\section{Scenario and System Model}
\label{scenario}


\subsection{Considered Scenario}

\begin{figure}
    \centering
    \includegraphics[width=0.97\linewidth]{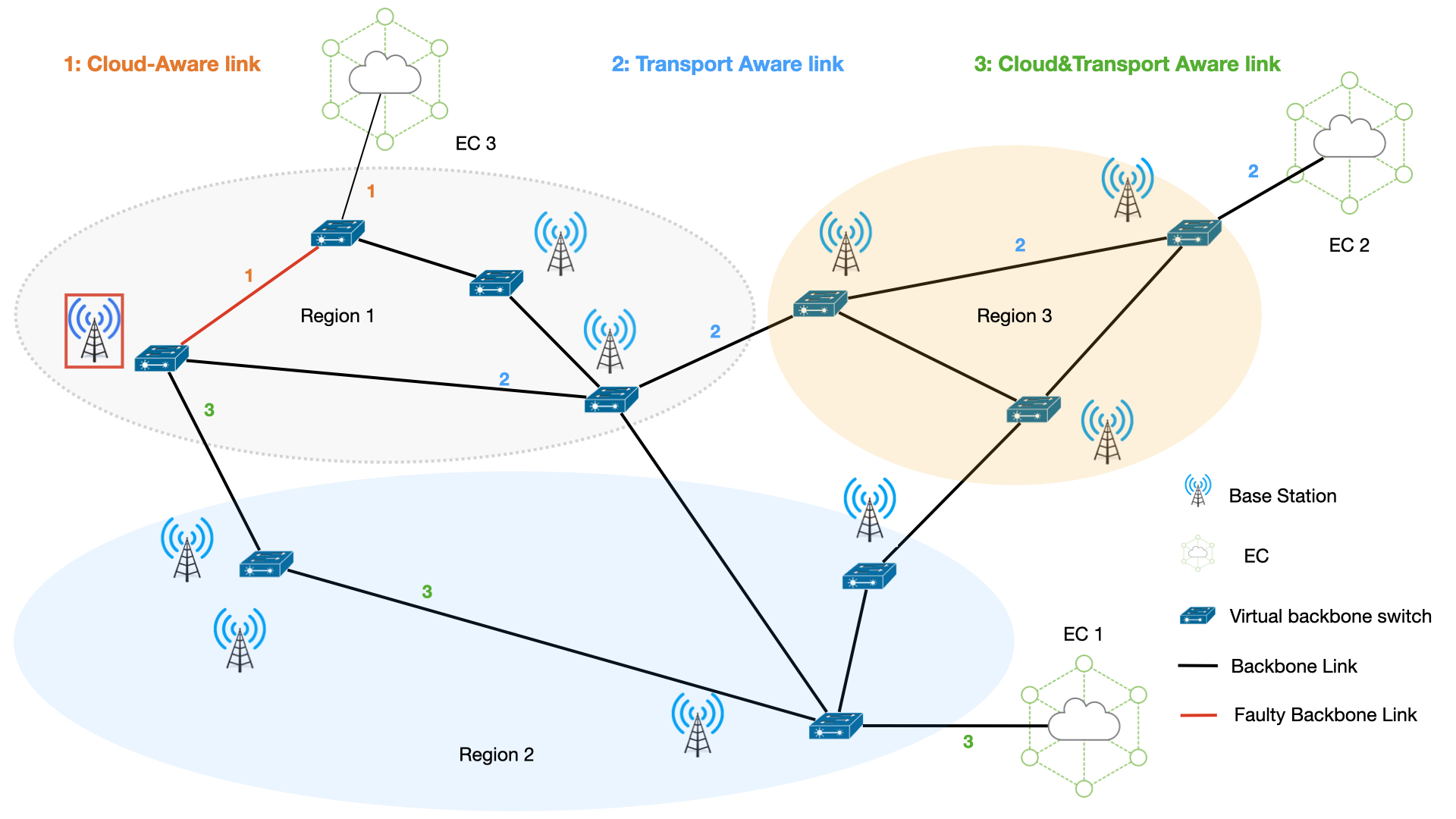}
    \caption{Edge computer nodes across different network regions}
    \label{fig:context}
\end{figure}

In this paper, we consider the joint use of cloud and transport KPIs to make the optimal decision in the placement of VNFs. Fig. \ref{fig:context} shows an example application scenario of our approach. If only cloud or transport parameters are considered separately during VNF placement decision making process (i.e. selecting the most appropriate EC), the requested service cannot be provided in some cases. For example, consider the utilization of link 1 (poor quality link) in Fig. \ref{fig:context} for only cloud aware decision making to reach $EC_3$ (EC with high cloud resource) and link 2 (good quality link) for only transport aware decision making to reach $EC_2$ (EC with low cloud resources). In both conditions, both transport and cloud level service requests cannot be provided efficiently. For this reason, the selection of a EC center must depend on both cloud and transport parameters as is done with link 3 (high quality link) to reach $EC_1$ (EC with high cloud resources). More specifically, in \cite{zeydan2021} the authors have demonstrated that considering both transport and cloud related parameters simultaneously, better EC decisions are taken by the previously trained ML models.

\vspace{-1mm}
\subsection{System Model and Problem Statement}
\label{sec_model}

In this paper, we make the following assumptions. The MNO operates a network with $\mathcal{K}=\{1,\ldots,K\}$ ECs, and provides a set $\mathcal{I}=\{1,\ldots, I\}$ of VNFs to users. Each user request asks for an instance of a VNF $i \in \mathcal{I}$. The requests for VNF type $i \in \mathcal{I}$ arrive and depart following Poisson processes with rate $\lambda_i$ and $\mu_i$, respectively.
The problem is assumed to be online - i.e. the demands arrive into the system one-by-one. The network management system keeps records of the VNFs served by each network EC node.  Each request $req$ is represented by a vector compounded of the number of CPU cores of the corresponding VNF and the bandwidth of the required traffic $[req_{cpu}, req_{bw}]$; e.g., $[3,100]$ is a request for a VNF that needs 3 CPU cores and processes 100 Mbps traffic. The requests are represented by the set $\mathcal{J} = \{req_1, req_2, ..., req_J\}$.


Under these assumptions, the problem is to place each given VNF request in a proper EC, without knowledge of future requests in order to maximize the acceptance rate. The solution to this problem should consider both the capacity constraints of the ECs and the bandwidth constraints of the links.


\vspace{-1mm}
\section{Optimal Solution}
\label{sec_MDP}
In this section we obtain the optimal solution of the problem by formulating it as a finite MDP then solving it using DP, i.e., the PI algorithm.

\subsection{The MDP Formulation}

An MDP \cite{10.5555/517430}, models controlled stochastic dynamical systems whose evolution is subject to random factors and which can be modified by certain decisions. An MDP provides a mathematical framework for learning sequential decision making, where actions in each state $s$  provide not only immediate rewards $\mathcal{R}$, but also the subsequent state $s'$. Mathematically, an MDP is a 5-tuple of $\langle \mathcal{S}, \mathcal{A}, \mathcal{P}, \mathcal{R}, s_0 \rangle$, where:

\begin{itemize}
    \item $\mathcal{S}$: the environment's finite set of states 
    \item $\mathcal{A}$: the finite set of applicable actions within the environment states
    \item $\mathcal{P}$: the state transition probabilities where\\ $\mathcal{P}(\mathcal{S}_{t+1}=s' \, | \, \mathcal{S}_t=s, A_t=a)$
    \item $\mathcal{R}(s, a, s')$: the immediate reward the agent obtains for being in state $s$, taking action $a$ and ending up in the subsequent state $s'$.
    \item $s_0$: the initial state at which the agent starts its task
\end{itemize}

We assume here that the environment is fully observable and the environment state at time $t$, denoted $s_t$, contains all the relevant environment information to the agent. It reflects not only the network status at time step $t$, but also the VNF $req_j$ arriving into the system at time $t$.

An MDP where all 5-tuple elements are known in advance, is called a model-based MDP and can be solved by DP methods, such as the PI algorithm. This allows to obtain the optimal agent's policy $\pi^*(s)$, in order for it to place as many VNFs as possible. 

We define the state $s_t$ as a vector of vectors $s_t = [[ecs], [d]]$, where $ecs$ represents as many vectors as there are EC nodes in the system, $ecs = [[M_1], [...], [M_K]]$. Each $M_k$ vector has size $I$ which is the total number of different VNF types. The value of the $i$-th element of $[M_k]$ determines the total number of active requests using instances of VNF type $i \in \mathcal{I}$, otherwise 0. Following the technique presented in \cite{bakhshi2021globe}, the vector $[d]$ has size $\mathcal{I}$ and only one of its $i$-th elements can take a value while the remaining elements are $0$; where $d[i] = +1$ if an incoming request $req$ requests  an instance of VNF type $i$, and $d[i] = -1$ if a request that have been using an instance of VNF type $i$ departs from the network.

The state transition probabilities $\mathcal{P}$ are crucial to compute the optimal agent’s policy $\pi^*(s)$. In this paper, the $\mathcal{P}$ are computed from the VNF $req$ arrival and departure rates, $\lambda_{i}$ and $\mu_{i}$ $\forall i \in \mathcal{I}$. To describe the process of computing $\mathcal{P}$ let us define two subsets of the state space, namely:

\begin{itemize}
    \item $\mathcal{S}^{+} \subset \mathcal{S} = \text{\{s} \, | \, \exists i \textbf{ s. t. } d_{i}=+1 \text{\}}$
    \item $\mathcal{S}^{-} \subset \mathcal{S} = \text{\{s} \, | \, \exists i \textbf{ s. t. } d_{i}=-1 \text{\}}$
\end{itemize}    



Whenever $s \in \mathcal{S}^+$, this implies a request $req_j$ that can either be served by a EC node chosen by the agent $\mathcal{A}(s) \in \{M_1, M_2, ..., M_K\}$ or rejected due to insufficient resources to handle it. The states $s \in \mathcal{S}^-$ are completely transparent to the agent, since it does not perform any action upon departure of requests; therefore, there is no agent action as such, we could say that the action is \textit{void}. Let us illustrate this with an example considering the following scenario: $\mathcal{I}=\{1,2\}$, $\lambda_1 = 2, \lambda_2 = 4$, and $\mu_1 = 0.25, \mu_2 = 1$.

\begin{figure}
    \centering
\includegraphics[width=0.97\linewidth]{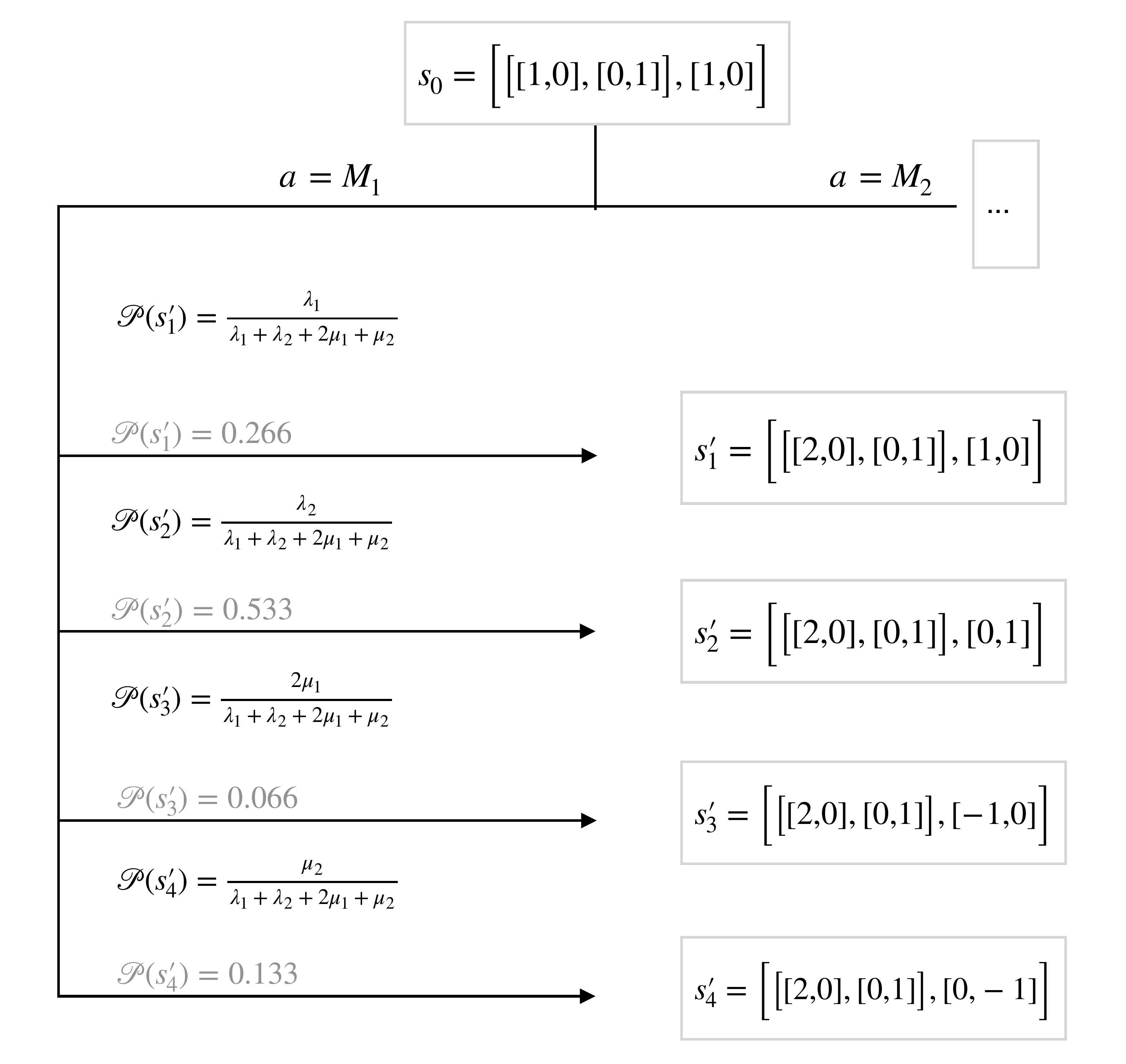}
    \caption{State transition probabilities for an arrival event}
    \label{fig:arrival_example}
\end{figure}

In Fig. \ref{fig:arrival_example}, $s_0$ reflects the network status where the type 1 request arrives at $t=0$. The agent chooses $a=M_1$, so is allocated on EC 1, which increments $M_1[1]$ by 1. After this action, there are several possible $s'$ that depend on the next event---a new $req$ of type 1 or 2 could arrive in the system, a departure of the $req$ of type 1 in $M_1$ or of type 2 in $M_2$ could also happen. Fig. \ref{fig:arrival_example} shows how the transition probabilities are calculated based on the transition rates using the competing exponentials theorem \cite{blumenfeld2001operations}. The numerator of $\mathcal{P}$ is the rate of the corresponding event, i.e., $\lambda_{i}$ for the states $s \in \mathcal{S}^{+}$, and for the departure states $s \in \mathcal{S}^{-}$, it is the total number of active $req$ of type $i$ that are held times the corresponding departure rate, $\mu_{i}$. The denominator is the total rate of all possible events in this state. Note that the arrival rates are independent and the departure rates depend on the total number of active $req$ of each type in the $M_k$. 

\begin{figure}
    \centering
\includegraphics[width=0.97\linewidth]{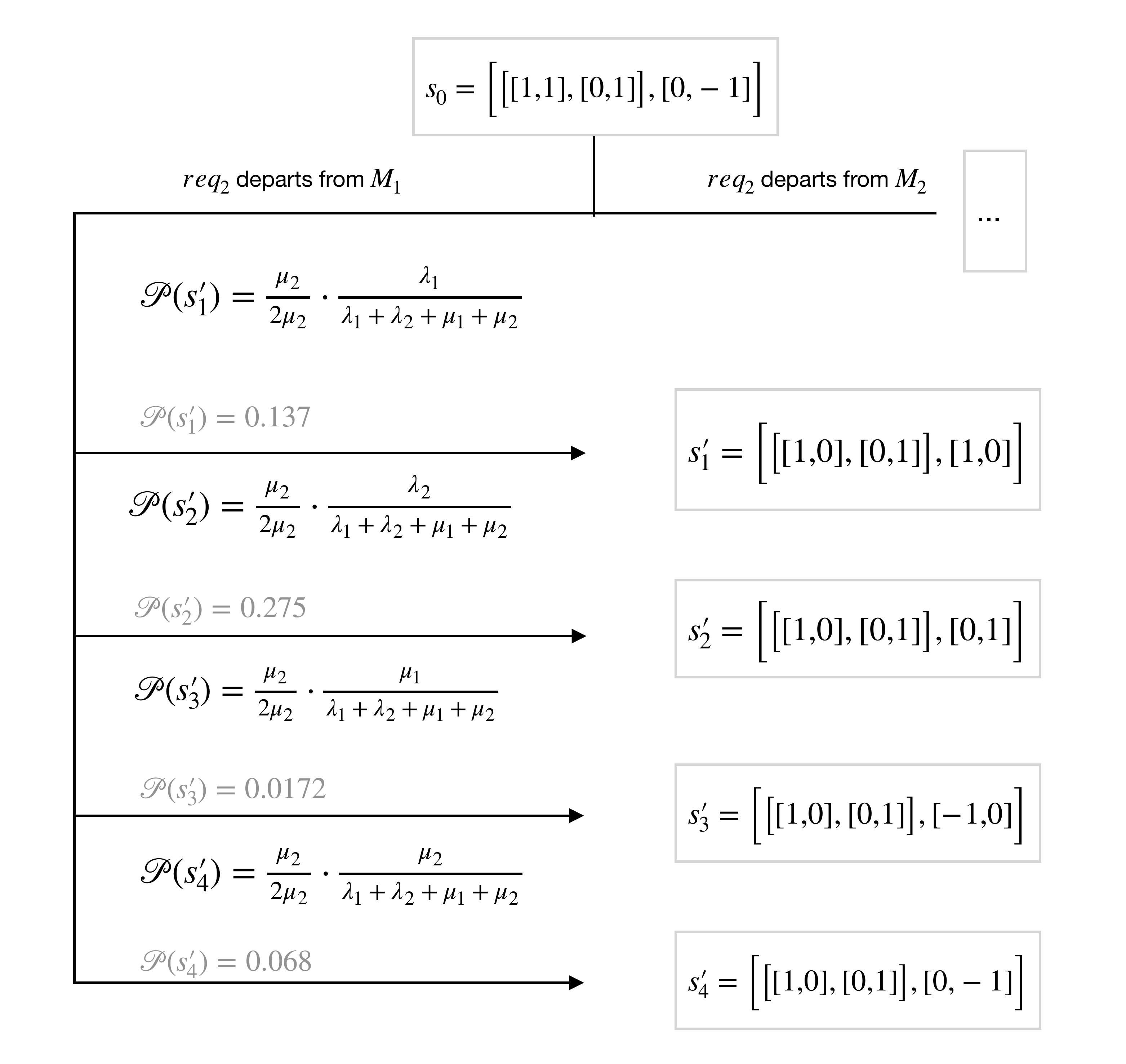}
    \caption{State transition probabilities for a departure event}
    \label{fig:departure_example}
\end{figure}

In Fig. \ref{fig:departure_example}, $s_0$ indicates that a $req$ of type 2 is departing from the system. It could be from $M_1$ or $M_2$. Fig. \ref{fig:departure_example} shows the state transition probabilities only for the case where the $req$ is departing from $M_1$; for $M_2$, it would be the same procedure as shown here. In this case, the $\mathcal{P}$ is composed of two terms, ${\mu_2}/{2\mu_2}$ is the probability of $req$ of type 2 that is departing from $M_1$, considering $\mu_2$ from $M_1$ and $M_2$. 
The second term, as shown in Fig. \ref{fig:arrival_example}, is obtained by the competing exponentials theorem. Note that the sum of probabilities $\mathcal{P}(s_i)$ is equal to $0.5$, since only half of the possible transitions are represented. In this paper, $\mathcal{R}(s, a, s') = 1$ if the given request is deployed in one of the EC nodes and $\mathcal{R}(s, a, s') = 0$  if the demands are rejected.


\vspace{-1mm}
\subsection{Dynamic Programming}
\label{sec_PI}

DP is a collection of methods by which it is possible to obtain the optimal policy of an MDP, as long as all elements of the model are known in advance. The PI algorithm is one of these  methods \cite{sutton2018}.

As shown in Algorithm \ref{alg_PI}, PI is divided into two sub-algorithms; $\textit{Policy Evaluation}$ and $\textit{Policy Improvement}$. The former computes $V_i(s)$ $\forall s \in \mathcal{S}$ for a given policy $\pi_i(s)$ . The latter improves the previously given policy $\pi_i(s)$ and obtains a new improved policy $\pi_{i+1}(s)$. 

The state-value function $V(s)$ pairs $s$ to $r$ and determines how good it is for the agent to be in a given $s$. The $V(s)$ can be expressed in terms of $\pi(s)$, where $V_{\pi}(s)$ describes how  good it was for the agent to follow its $\pi(s)$ in a particular $s$, take an $a$ and transition to another $s'$. The immediate $\mathcal{R}(s,a,s')$ plus the function $V(s')$  of the landed next state $s'$ determines how good the original $s$ was; more precisely, we have 

\begin{equation*}
V_{\pi}(s) = \sum_{a} \pi(a \, | \, s) \sum_{s'}  \mathcal{P}[\mathcal{R}(s,a,s') + \gamma V(s')]
\end{equation*}

The discount rate $\gamma \in [0, 1]$ prevents the agent from infinitely returning to a state accumulating rewards. If $\gamma \approx 1$ the agent will prioritize expected future rewards. In contrast, when $\gamma \approx 0$ the agent will strongly consider the immediate rewards. Each iteration is guaranteed to result in an improved new policy until the optimal strategy is obtained. Since a finite MDP is a finite set of $\mathcal{S}$ and $\mathcal{A}$, convergence of $V^*(s)$ to obtain the $\pi^*(s)$ is achievable in a finite number of iterations. Prior knowledge of all valid $s$, allows to set $\gamma \approx 1$.

Finally, out of all possible policies, there's at least one to be better or equally as good to all other policies achieving the optimal state-value function $V^*(s)$; more specifically:

\begin{equation*}
V^*(s) \gets \max_{\pi} V_{\pi}(s)    \ \ \ \ \forall s\in \mathcal{S}.
\end{equation*}

\begin{algorithm}
\caption{$\textit{Policy Iteration}$}
\label{alg_PI}
\begin{small}
\begin{algorithmic}
	\State Randomly initialize $V(s) \in \mathbb{R}$ and $\pi(s) \in {\mathcal{A}}(s)$   $\forall s \in {\mathcal{S}}$
   	\While {$\Delta > \theta$} \Comment{The $\textit{Policy Evaluation}$ loop ~~~~~~~~~~~~~~$\,$}
   	    \State \Comment{$\theta$ determines the accuracy of estimation}
		\State $\Delta \gets 0$
		\For {$\forall s \in \mathcal{\mathcal{S}}$}
			\State $v \gets V(s)$
			\State $V(s) \gets \sum_{s', r} p(s',r \, | \, s,\pi(s))[r+\gamma V(s')]$
			\State $\Delta \gets \text{max}(\Delta, |v - V(s)|)$
		\EndFor
	\EndWhile
		
	\State $policy_{stable} \gets true$
	\For {$\forall s \in \mathcal{S}$} 
	    \Comment{The $\textit{Policy Improvement}$ loop}
		\State $action_{old} \gets \pi(s)$
		\State $\pi(s) \gets \text{argmax}_{a} \sum_{s', r} p(s',r \, | \, s,a)[r+\gamma V(s')]$
		\If {$action_{old} \ne \pi(s)$} 
	        \State$policy_{stable} \gets false$
		\EndIf
		\EndFor
	\If {$policy_{stable}$} 
	    \State \Return $\pi^*$
	\Else
	    \State $\text{go to $\textit{Policy Evaluation}$ loop}$
	\EndIf
\end{algorithmic}
\end{small}	
\end{algorithm}

\vspace{-1mm}
\section{Practical Solutions 
}
\label{sec_QLRL}

In the MDP formulation, we defined a subset $\mathcal{S}^{-}$ to derive the transition probabilities in the case of departure from a demand, as in Fig. \ref{fig:departure_example}. Since it is not needed in the practical solution, we redefine the state as follows: 

\begin{equation*}
s = [req_{cpu}, req_{bw}, M_{1}^{fcpu}, ..., M_{K}^{fcpu}, M_1^{fbw}, ..., M_K^{fbw}] 
\end{equation*}

where $M_k^{fcpu}$ is the available CPU units of the $k$-th EC node, and $M_k^{fbw}$ is the available BW connection to reach the $k$-th EC node.

\subsection{{Q-Learning}}

In cases where not all aspects of the system are known in advance, model-free reinforcement learning approaches can find the (near) optimal policy. Here, the agent must learn from its own actions without a given $\pi(s)$; hence, it is also an off-policy model.

The \textit{Q-Learning} algorithm \cite{Watkins:1989} drives the agent's learning by assigning values to $(s, a)$ pairs. 
The Q-values define how good an action is in a given state. They are updated for  each interaction with the environment by the following rule:

\begin{equation*}
\begin{split}
\underbrace {Q(s,a)}_{\text{new value}}  \gets \underbrace {Q(s,a)}_{\text{old value}} + \underbrace {\alpha }_{\text{learning rate}}\cdot
\ \ \ \ \ \ \ \ \ \ \  \ \ \ \ \ \ \ \  \ \ \ \ \ \ \ \ \ \ \ \ \ \ \ \\\
\overbrace {{\bigg (}\underbrace {\underbrace {\mathcal{R}(s, a, s')} _{\text{reward}}+\underbrace {\gamma } _{\text{discount rate}}\cdot \underbrace {\max _{a'}Q(s',a')} _{\text{estimate of optimal future value}}} _{\text{new value (temporal difference target)}}-\underbrace {Q(s,a)}_{\text{old value}}{\bigg )}} ^{\text{temporal difference}}
\end{split}
\end{equation*}

By introducing a learning rate  $\alpha \in [0, 1]$, it is possible to control the variation of Q-values, where $\alpha$ defines to what degree the agent replaces the old data with new ones. A rate $\alpha \approx 1$ forces the agent to consider the latest information, while $\alpha \approx 0$ causes the agent to learn nothing.  In the  update rule of Q-Learning, the new Q-value is the weighted combination between the old Q-value and the new observation that the agent must believe. \textit{Q-Learning} algorithm converges to an optimal Q-value, $Q^*(s,a)$, given sufficient $\alpha$ and exploration over $\mathcal{S}$ that satisfies the Robbins-Monro conditions \cite{stochasticaprox} \cite{qlearninghistory}.

A disadvantage of \textit{Q-Learning} is that the agent can only learn from actions performed in visited states, otherwise there is no learning. The chain of successive actions and the resulting states form an episode. This leads to the exploitation/exploration dilemma. It might be interesting for the agent, especially at the beginning of its training\footnote{There is no training concept as such in \textit{Q-Learning} as is exists in other ML domains. Nonetheless, we refer to training as the initial episodes in which the agent attempts to populate its Q-Table.},  adopt an explorer profile to visit as many states and try as many actions as possible within those states. When the agent's learning reaches a certain level, it is more beneficial to exploit the known actions that bring higher rewards when revisiting the known states. This strategy is known in the literature as $\epsilon$-greedy \cite{sutton2018}. 

\textit{Q-Learning} differs from PI in that it cannot be evaluated directly without the agent having prior experience. That is, it requires knowledge about what to do when confronted with an $s$, otherwise the agent's actions would be completely random according to the $\epsilon$-greedy strategy. Therefore, the \textit{Q-Learning} agent is trained with several different VNF $req$ sequences before using its agent to evaluate. To prevent \textit{Q-Learning} agent from being confronted with an unknown state (during the evaluation stage) and failing its attempt to search for it in Q-Table, the algorithm has been modified to allow it to learn during the evaluation stage as well.

\begin{algorithm}
	\caption{\textit{Q-Learning}}
	\label{alg_QL}
	\begin{small}
\begin{algorithmic}
    \State \text{Set values for: }\text{learning rate} $\alpha$, \text{discount rate} $\gamma$
	\State Randomly initialize $Q[s,a] \in \mathbb{R}$  $\forall s \in {\mathcal{S}}$, $\forall a \in {\mathcal{A}}(s)$
	\For {each episode}
		\State Initialize $\mathcal{S}$
		\For {each step}
		    \If{evaluation}
			    \If{$s$ in $Qtable$}
			        \State $a \gets $ $argmax(Q(s, a))$
			        
			    \Else
			        \State behave as in  $not$ evaluation
			\EndIf
			\Else
			    \State $a \gets $ action from ${\mathcal{A}}(s)$ by $\epsilon$-greedy strategy
	    	    \State Observe $s'$
	            \State $r \gets \mathcal{R}(s,a,s')$
		        \State $Q(s,a) \gets Q(s,a) + \alpha [ r + \gamma \max_{a'}Q(s',a') - Q(s,a)]$
        	    \State $s \gets s'$
            \EndIf
		\EndFor
		\State Until $S_T$
	\EndFor
\end{algorithmic}
\end{small}	
\end{algorithm}

Another important aspect to consider in the present work is that there is no absorbing/terminal state $S_T$ in the VNF placement problem. In episodic tasks, a $S_T$ is reached when the agent reaches its goal or commits an error that forces the environment to restart. However, in our problem, the VNFs assignment  should be done as long as there are requests. Therefore, in this work $S_T$ is determined by the total number of $req_j$ in the VNF $req$ training files.

\subsection{{Best Fit}}

In addition to Q-Learning, we develop another practical solution called \textit{best fit} which is inspired by a classical Weighted Round Robin load balancing algorithm. The algorithm assigns the incoming VNF request to the EC that has the highest network metric value $l$ which is defined as follows,

\begin{equation*}
l= a\ \frac{M_k^{fcpu}}{100} + (1-a) \ \frac{1}{Num_{hops}}
\end{equation*}
where $Num_{hops}$ is the number of hops from the BS to the EC node, and $a = {M_k^{ubw}}\ /\ ({\text {Total Network BW})}$ where $M_k^{ubw}$ is the BW connection used to reach the $k$-th EC node.

The algorithm \textit{best fit} first checks if all EC nodes have enough CPU and BW resources, otherwise a rejection is generated. If there is only one EC with available resources, the VNF $req$ is assigned to that EC node. If there is more than one EC node with the same resource availability,  the VNF $req$ in this paper is randomly assigned. If there is more than one EC node with different  resources availability, the VNF $req$ is assigned to the  EC node that is proved to have the highest $l$.

\vspace{-1mm}
\section{Simulation Results}
\label{sec_sim}


The following simulations were performed from a theoretical perspective, since the PI algorithm requires a large amount of computational resources and execution time. It would be a tedious task to simulate a network with a large number of entities. For this reason, several BSs demanding $req$ and two EC nodes were considered to meet the demands. The goal is to measure the performance of \textit{Q-Learning} and observe how far it is from a mathematical solution, a model-based MDP, and how much it can outperform \textit{best fit}. During the evaluation stage all three algorithms are run for only one episode over the same set of VNF requests. Unless otherwise specified, the following simulations are run with the settings defined in Table \ref{table:default_sim_settings}.

\begin{table}[t]
\centering
\caption{{Default Simulation Settings}}
\vspace{-2mm}
\begin{small}
\begin{tabular}{|c c|}
 \hline
 \textbf{Settings} & \textbf{Values} \\ [0.1ex] 
 \hline\hline
 Initial network status & $[4, 12, 0, 0, 1000, 400, 0, 0]$  \\ 
 \hline
 No. training files & $10$  \\
 \hline
 No. $req$ in training files & $500$  \\
 \hline
 No. episodes in training files & $250$  \\
 \hline
 No. evaluation files & $20$  \\ 
 \hline
 No. $req$ in evaluation files& $500$\\
 \hline
 [$\lambda_1, \lambda_2]$&$[3, 2]$\\
 \hline
 [$\mu_1, \mu_2]$&$[1, 0.5]$\\
 \hline
 [$req_1, req_2]$&$[(1, 300), (3, 50)]$\\
 \hline
\end{tabular}
\label{table:default_sim_settings}
\end{small}
\end{table}

\subsection{Influence of $\alpha$ and $\gamma$ in \textit{Q-Learning} algorithm}

Continuous tasks, such as the one considered here, force the agent to compromise by achieving a high reward in the long-run but giving enough importance to each current state value. The agent must learn to some degree, but without constantly overriding  what it has already learned. Therefore, $\alpha$ and $\gamma$ must be configured considering the nature of the task the agent is expected to develop.

\begin{figure}
  \subfloat[$\alpha = 0, \gamma = 0.001$]{
	 \centering
	 \hspace{-3mm}
	 \includegraphics[width=0.49\linewidth]{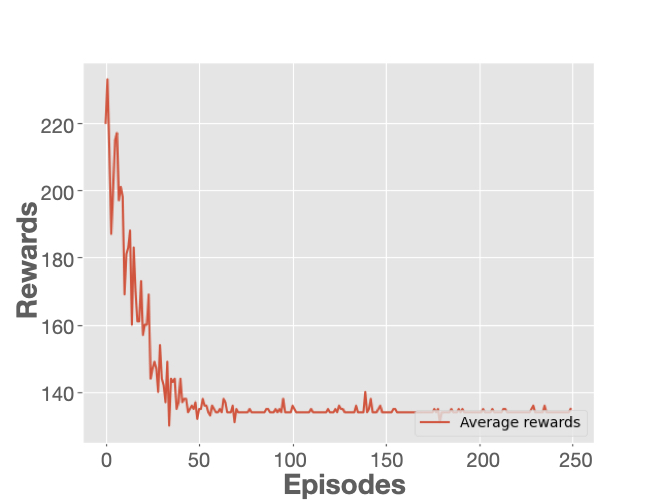}
  }
  \hfill
  \subfloat[$\alpha = 0.5, \gamma = 0.5$]{
	 \centering
     \hspace{-3mm}
	 \includegraphics[width=0.49\linewidth]{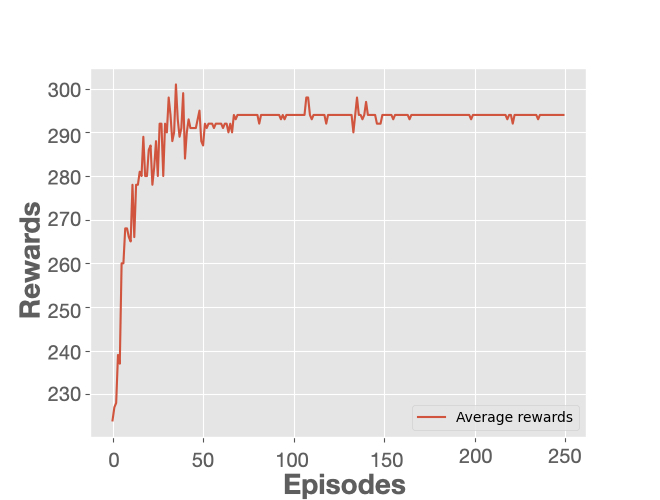}
  }

\vspace{-2mm}
  \subfloat[$\alpha = 0.9, \gamma = 0.9$]{
	 \centering
	 \hspace{-3mm}
	 \includegraphics[width=0.49\linewidth]{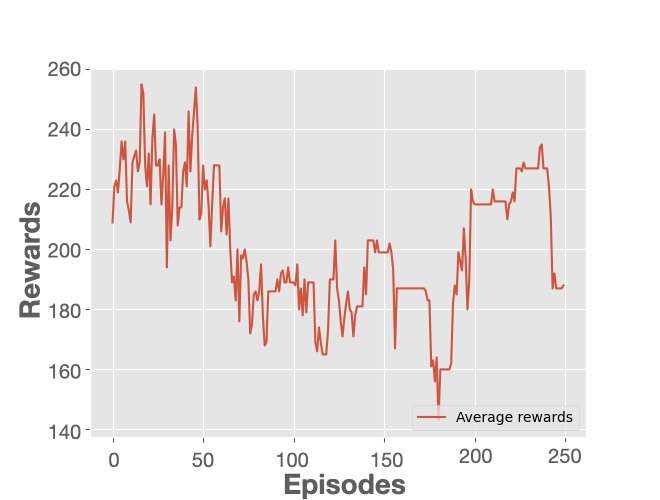}
   }
  \hfill	
  \subfloat[$\alpha = 0.5, \gamma = 0.9$]{
	 \centering
	 \hspace{-3mm}
	 \includegraphics[width=0.49\linewidth]{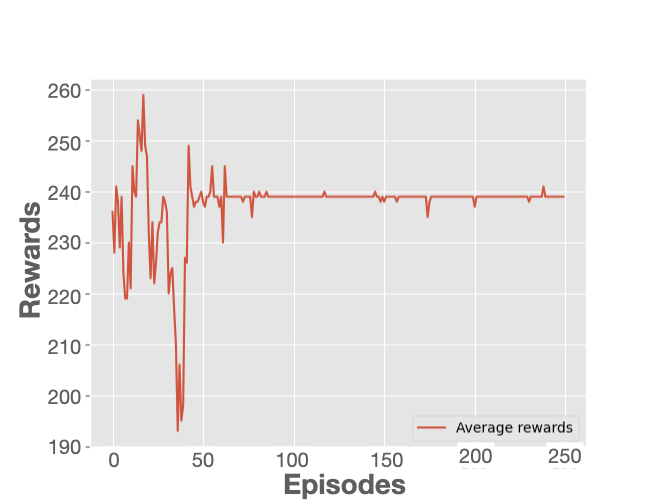}
  }
\caption{Influence of $\alpha$ and $\gamma$ in \textit{Q-Learning} algorithm}
\label{fig:alpha_gamma}
\end{figure}

Fig. \ref{fig:alpha_gamma} shows four different \textit{Q-Learning} simulations representing the average reward (Y-axis) collected by the agent over the course of 250 episodes (X-axis) through the same 500 VNF $req$ sequence file. Note that setting $\alpha$ and $\gamma$ to consider the most recent information and to favor the long-term reward does not ensure the desired learning convergence, Fig. \ref{fig:alpha_gamma}(c). This is mainly because there is no $S_T$ that determines the goal to be achieved. Setting $\alpha$ and $\gamma$ to $0.5$ have been show to be satisfactory.

\subsection{Influence of $\epsilon$-greedy parameters in \textit{Q-Learning} algorithm}
 
 \begin{figure}
  \subfloat[$\epsilon_{decay}= 0.1$]{
	 \centering
	 \hspace{-3mm}
	 \includegraphics[width=0.49\linewidth]{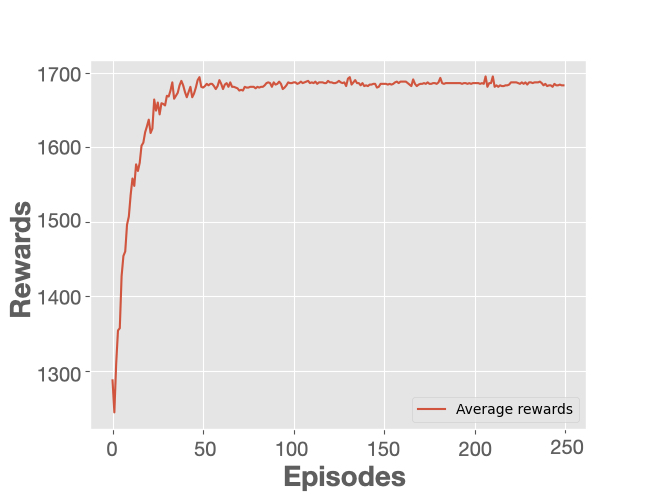}
   }
 \hfill 	
  \subfloat[$\epsilon_{decay}= 0.01$]{
	 \centering
	 \hspace{-3mm}
	 \includegraphics[width=0.49\linewidth]{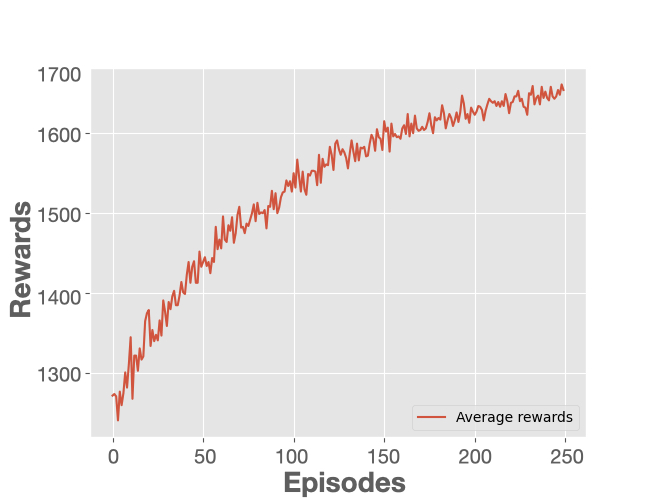}
    }

\vspace{-2mm}
  \subfloat[$\epsilon_{decay}= 0.001$]{
	 \centering
	 \hspace{-3mm}
	 \includegraphics[width=0.49\linewidth]{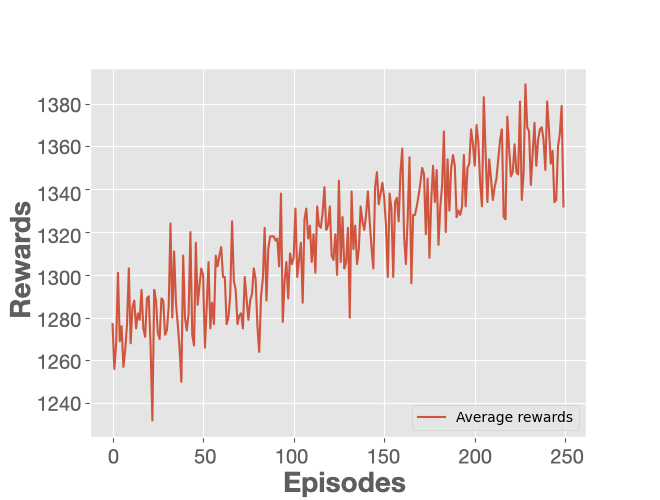}
    }
 \hfill	
  \subfloat[$\epsilon_{decay}= 0.03$]{
	 \centering
	 \hspace{-3mm}
	 \includegraphics[width=0.49\linewidth]{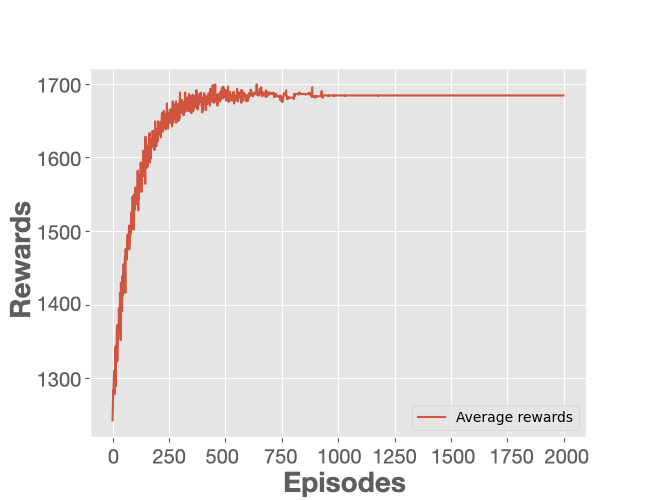}
    }
\caption{Influence of $\epsilon$-greedy parameters in \textit{Q-Learning} algorithm}
\label{fig:epsilon}
\end{figure}

The \textit{Q-Learning} $\epsilon$-greedy sub-algorithm that determines how much of an explorer or exploiter the agent is has certain parameters that define exactly how long the agent will take on such profiles. The decay rate ($\epsilon_{decay}$) has huge impact on the  convergence of the agent’s learning performance, as shown in Fig. \ref{fig:epsilon}. 

Depending on the number of episodes, a very small $\epsilon_{decay}$ rate may mean that the agent can never exploit what it has already learned, as in Fig. \ref{fig:epsilon}(c). By leaving a minimum epsilon value $\epsilon_{min} = 0.001$  and a $\epsilon_{max}=1$, the probability that the agent explores from time to time increases, even towards the end of the training stage. This is the reason why certain peaks are seen along the performance curve once it has converged Fig. \ref{fig:epsilon}(a). Fig. \ref{fig:epsilon}(d) shows the same level of convergence and the same average reward as in Fig. 5 (a) but over the course of 2000 episodes instead of 250 with an $\epsilon_{decay} = 0.03$. $\epsilon$ decays after each episode using: $\epsilon = \epsilon_{min} + (\epsilon_{max} - \epsilon_{min}) \cdot e^{(-\epsilon_{decay} \cdot episode_i)}$.

\subsection{Rejection ratio with respect to VNF arrival rate}

\begin{figure}[t]
    \centering
    \hspace*{-0.4cm}
    \includegraphics[width=0.97\linewidth]{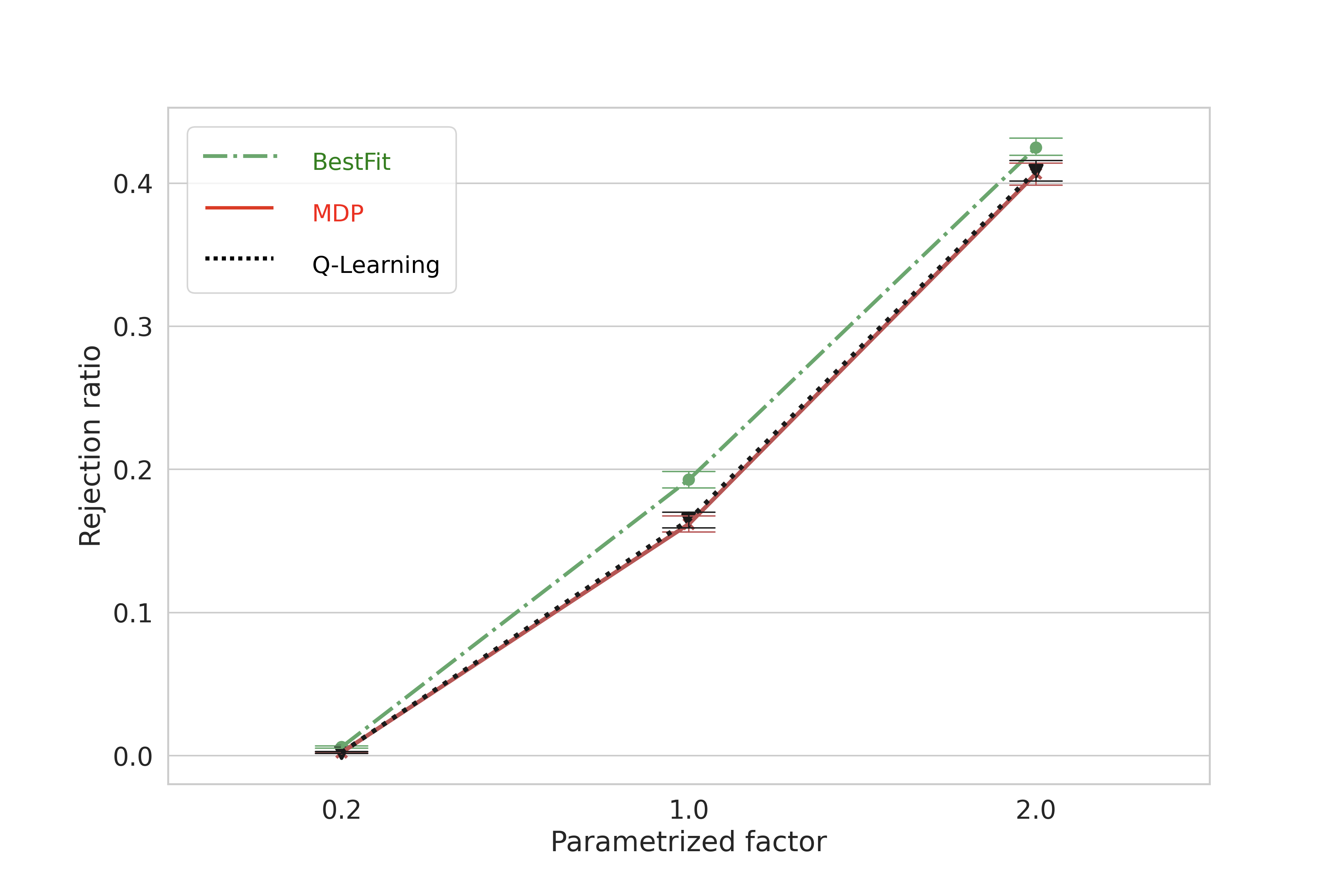}
    \vspace{-3mm}
    \caption{Rejection ratio with respect to VNF arrival rate}
    \label{fig:rate_results}
    \vspace{1mm}
\end{figure}

In these simulations, the performance of all three algorithms is evaluated using different $\lambda_i$ rates, as shown in Table \ref{table:rate}. The goal is to analyze how the $req_j$ arrival rate affects the resources of the EC nodes. The agent is forced to handle consecutive incoming requests with short and high arrival rate. The results are shown in Fig. \ref{fig:rate_results}, where the X-axis represents the different factor values by which the VNF arrival rates are multiplied.

In Simulation 1, the short $\lambda_2$ and $\mu_2$ ensure that such a demand arrives and departs in a short time, returning the resources to the EC center to which they were allocated, and thus becoming available again. As $req_j$ inter-arrival times increases, (Simulations 2 and 3), the MDP achieves a slightly better policy than \textit{Q-Learning}. The agent must constantly learn  how to assign the demands among the resources of EC nodes  as they are occupied for longer time steps. This occasionally results in rejections. Nevertheless, the performance difference is very small, indicating that the agent behaves optimally when EC centers' resources are scarce.

\begin{table}[t]
\centering
\caption{{Simulation Settings to Analyze Arrival Rate}}
\vspace{-2mm}
\begin{small}
\begin{tabular}{|c c c c|}
 \hline
 \textbf{Parameter} & \textbf{Sim. 1} & \textbf{Sim. 2} & \textbf{Sim. 3}\\ [0.1ex] 
 \hline\hline
 
 Factor & 0.2 & 1.0 & 2.0  \\
 \hline
 [$\lambda_1, \lambda_2]$& [0.6, 0.4]&[3, 2]&[6, 4]   \\ 
 \hline
 [$\mu_1, \mu_2]$&$[1, 0.5]$&$[1, 0.5]$&$[1, 0.5]$\\
 \hline
\end{tabular}
\label{table:rate}
\end{small}
\end{table}

\subsection{Rejection ratio with respect to EC and link capacity}

\begin{table}[t]
\centering
\caption{{Simulation Settings to Analyze EC Resources}}
\vspace{-2mm}
\begin{small}
\begin{tabular}{|c c c c|}
 \hline
 \textbf{Parameter} & \textbf{Sim. 1} & \textbf{Sim. 2} & \textbf{Sim. 3}\\ [0.1ex] 
 \hline\hline
 
 Factor & 0.8 & 1.0 & 1.2  \\
 \hline
 
 $[M_1^{fcpu}, M_2^{fcpu}]$ &[4, 8] &[5, 10] &[6, 12]\\
 \hline
 $[M_1^{fbw}, M_2^{fbw}]$  &[800, 320] &[1000, 400] &[1200, 480]\\
 \hline
\end{tabular}
\label{table:capacity}
\end{small}
\end{table}

\begin{figure}
    \centering
    \hspace*{-0.4cm}
    \includegraphics[width=0.97\linewidth]{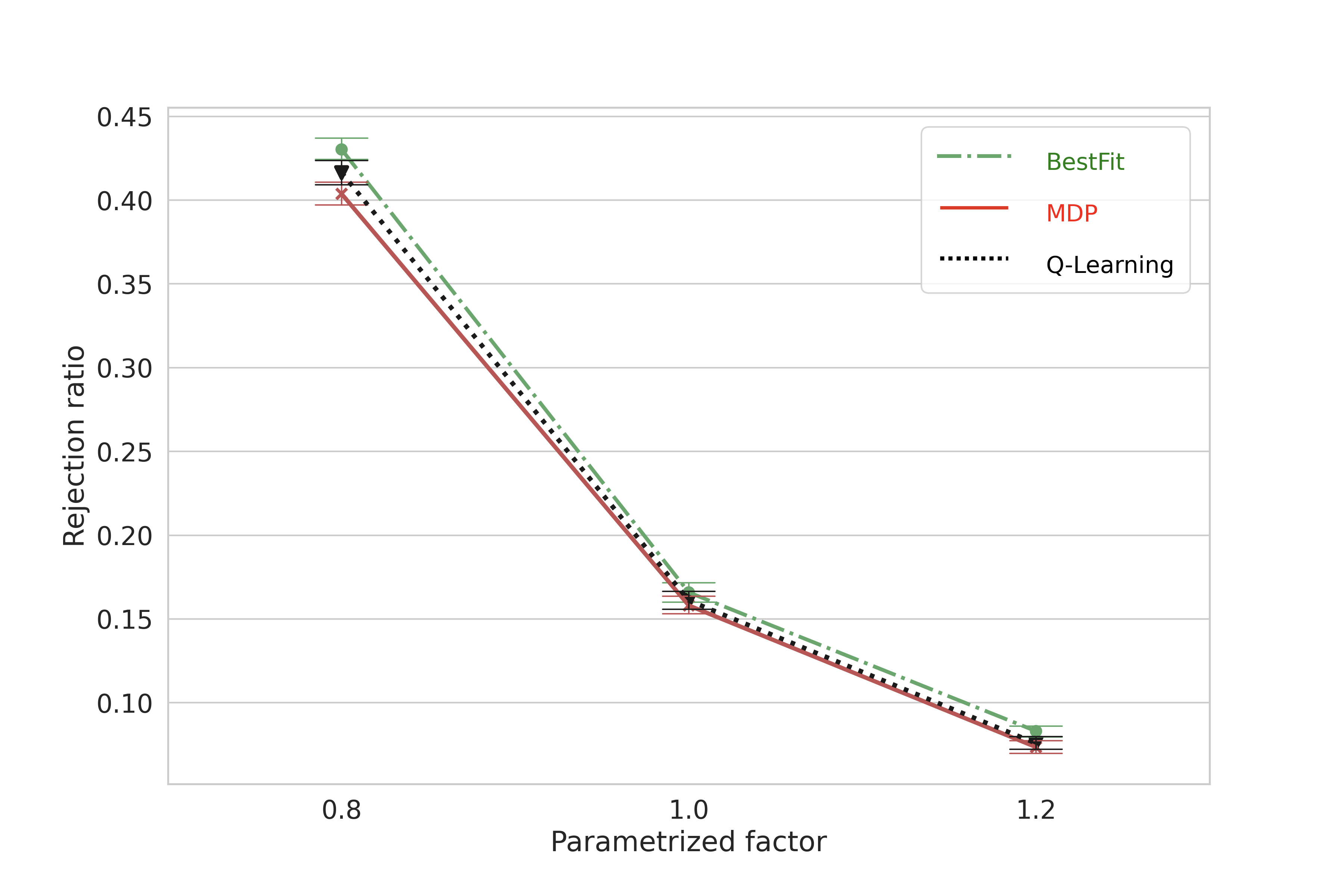}
    \vspace{-3mm}
    \caption{Rejection ratio with respect to EC and link capacity}
    \label{fig:capacity_result}
    \vspace{1mm}
\end{figure}

In these simulations, we evaluate the performance of the algorithms  with different CPU cores and BW link capacities, as shown in Table \ref{table:capacity}. The results are shown in Fig. \ref{fig:capacity_result}.

In Fig. \ref{fig:capacity_result}, for Simulation 1, the scarce available resources, for either $M_1^{fcpu}$ or $M_2^{fbw}$, determine the performance difference between MDP and \textit{Q-Learning}. The \textit{Q-Learning}'s policy is better than that of \textit{best fit}, but worse than that of MDP. The $M_1^{fcpu}$ are used up very quickly, so the rest of the outcome is somehow deterministic. As the available resources are increased in each EC, \textit{Q-Learning} appears to improve its learning and to closely converge to ideal policy of the MDP. It should be noted that the randomness of \textit{Q-Learning} during the exploration stage, determines, to some extent, the starting point for building the path for the optimal policy. When the exploitation stage begins, the probabilities of changing what is already known are less likely.

\subsection{Rejection ratio with respect to EC resource heterogeneity}

\begin{figure}[t]
    \centering
    \hspace*{-0.4cm}
    \includegraphics[width=0.97\linewidth]{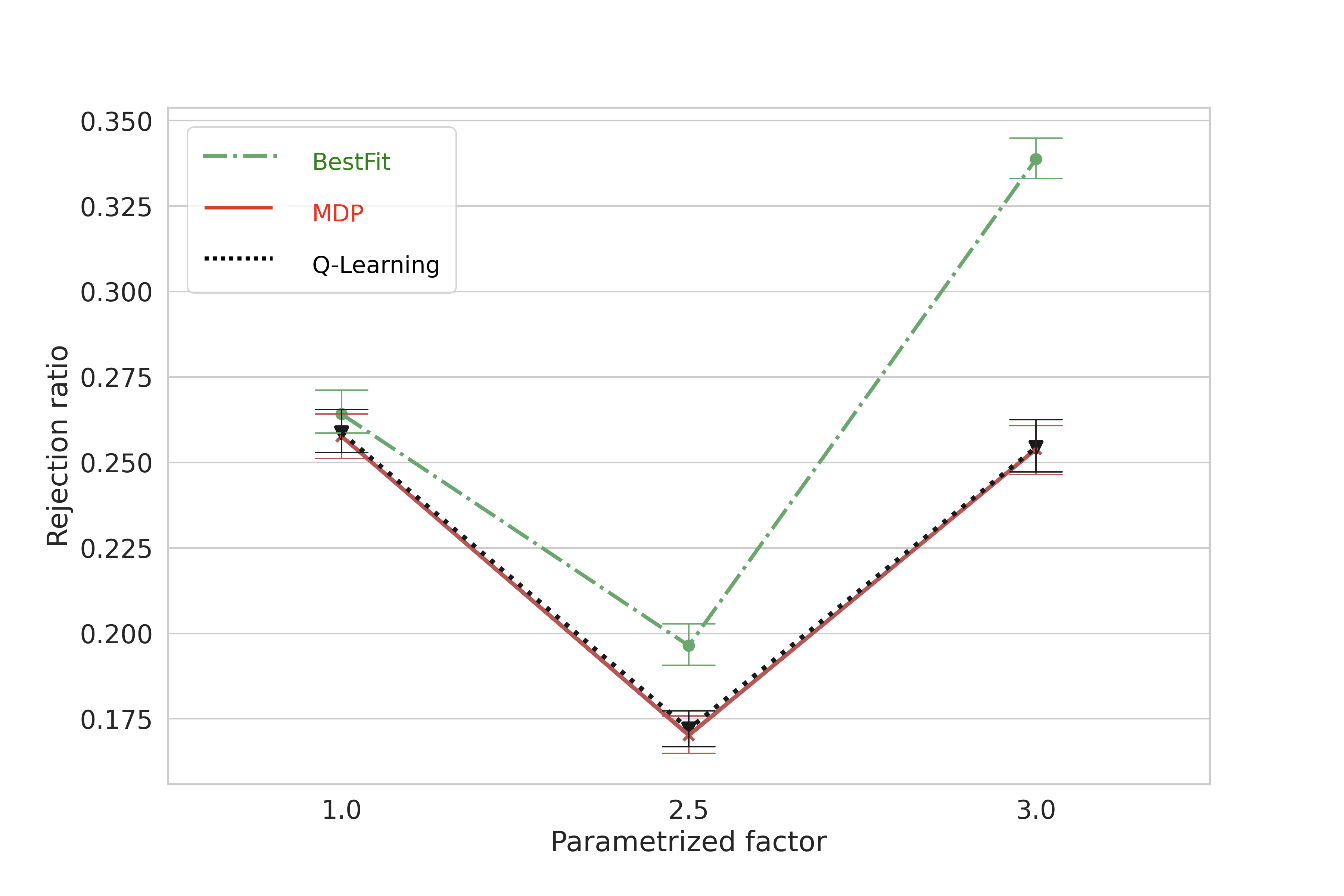}
    \vspace{-3mm}
    \caption{Rejection ratio with respect to EC resource heterogeneity}
    \label{fig:mec_het_result}
    \vspace{1mm}
\end{figure}

In this simulations, we analyze how the agent can effectively reflect the heterogeneity of EC resources. To this end, 
in  simulations EC 1's available resources are fixed, i.e. $[4, 1000]$, while EC 2's CPU cores are increased as $EC_2^{cpu}=\beta \cdot EC_1^{cpu}$, and its available BW is decreased as $EC_2^{BW}=\frac{1}{\beta} \cdot EC_1^{BW}$. The simulation settings and corresponding results are  shown in Table \ref{table:mec_het} and Fig. \ref{fig:mec_het_result} respectively.

The first thing that can be observed in all simulations is the small influence of each resource parameter on the agent's decision. It is clear that the agent treats the lack of resources similarly whether it is CPU or BW. It is noticeable that simulation 1 and 3 show almost the same results in terms of rejection ratio, and the difference between the two simulations is a factor 3. Simulation 2 also shows the same dynamics as all previous simulations: Q-Learning performs slightly worse when the resources are larger and there are more opportunities to assign $req_j$ among the network's EC centers.

\begin{table}[t]
\centering
\caption{{Simulation Settings to Analyze EC Heterogeneity}}
\vspace{-2mm}
\begin{small}
\begin{tabular}{|c c c c|}
 \hline
 \textbf{Parameter} & \textbf{Sim. 1} & \textbf{Sim. 2} & \textbf{Sim. 3}\\ [0.1ex] 
 \hline\hline

 $\beta$ & 1.0 & 2.5 & 3.0\\
 \hline
 $M_2^{fcpu}$ &[4] &[10] &[12]\\
 \hline
 $M_2^{fbw}$ &[1000] &[400] &[333]\\
 \hline
\end{tabular}
\label{table:mec_het}
\end{small}
\end{table}

\subsection{Rejection ratio with respect to VNF demand heterogeneity}

\begin{table}[t]
\centering
\caption{$req_2$ Demand Settings to Analyze Demand Heterogeneity}
\vspace{-2mm}
\begin{small}
\begin{tabular}{|c c c c|}
 \hline
 \textbf{Parameter} & \textbf{Sim. 1} & \textbf{Sim. 2} & \textbf{Sim. 3}\\ [0.1ex] 
 \hline\hline
 $\beta$ &1.5 &2 &4\\
 \hline
 $req_2^{CPU}$ & 3 & 4  & 8 \\
 \hline
 $req_2^{BW}$ &  200  &  150  &  75 \\
 \hline
\end{tabular}
\label{table:demand_het}
\end{small}
\end{table}

In these simulations, we proceed similarly to the previous one, but with implications for the VNF demand values. In this case, the $req_1$ demands remain static and the $req_2$ values are changed according to the Table \ref{table:demand_het}. 

\begin{figure}[t]
    \centering
    \hspace*{-0.4cm}
    \includegraphics[width=0.97\linewidth]{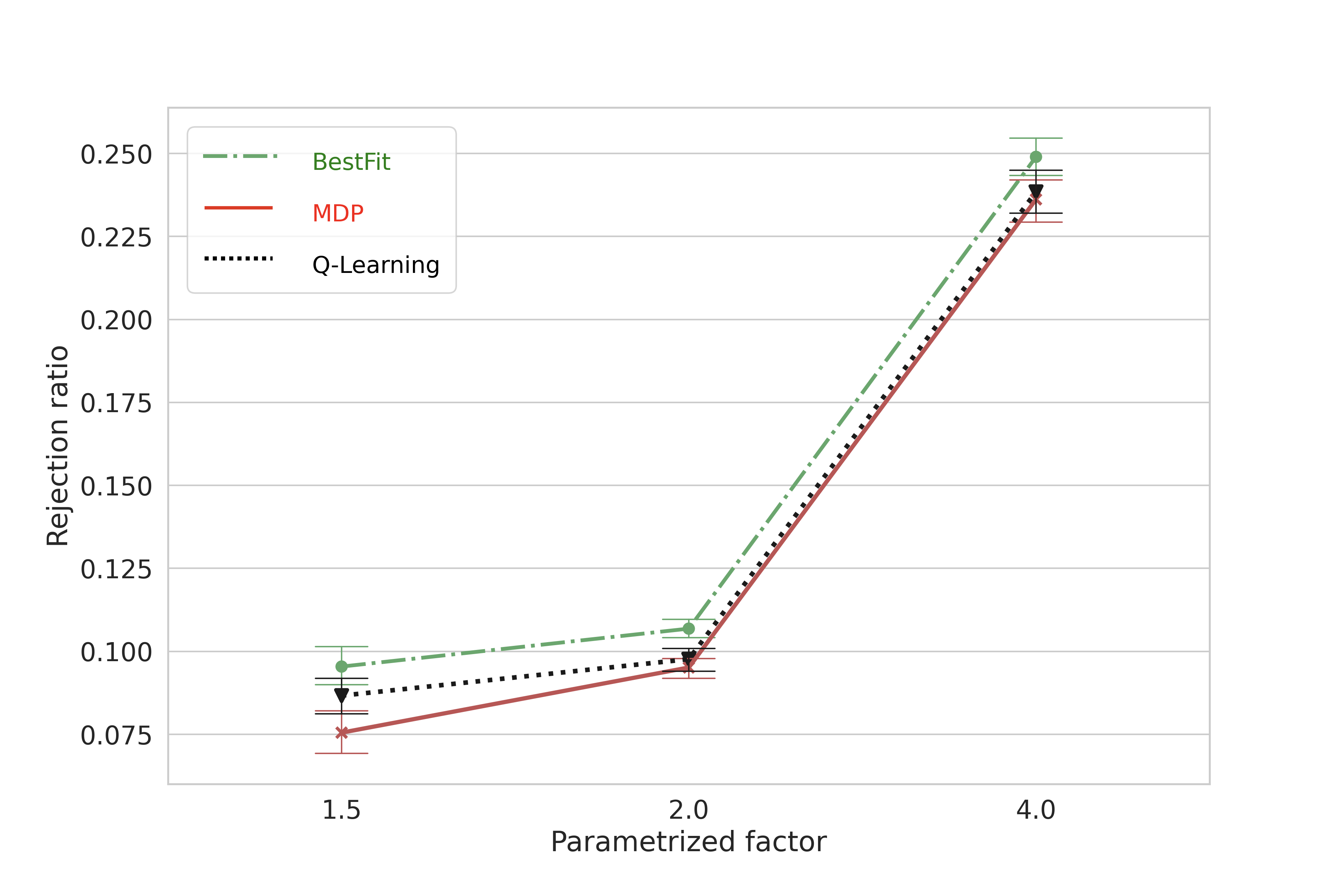}
    \vspace{-3mm}
    \caption{Rejection ratio with respect to VNF demand heterogeneity}
    \label{fig:demand_het_result}
    \vspace{1mm}
\end{figure}

For $req_2$, as the number of CPU requests increases, $req_{2cpu}=\beta \cdot req_{1cpu}$, the required link BW is lowered, $req_{2BW} = \frac{1}{\beta} \cdot req_{1BW}$. The results are shown in Fig. \ref{fig:demand_het_result}.  In Fig. \ref{fig:demand_het_result}, for $\beta=1.5$, a similar agent behavior, when available resources are not critical, can be seen. Seems to perform worse than MDP. As EC resources begin to be scarce, \textit{Q-Learning} approximates the MDP, resulting in similar policies.

\section{Conclusions and Future Work}

In this paper, we have studied the problem of VNF placement in EC-enabled 6G networks, considering both computational and communication resources. We have developed theoretically optimal and practical solutions to this problem. We obtained the former  by formulating the problem as a finite MDP, solved via PI. The latter is the model-free reinforcement learning approach. We evaluated the solutions in a wide range of network parameter settings. It has been shown that there is a striking performance similarity between Q-Learning and PI, especially when both algorithms face limited EC resources. Nevertheless, the MDP needs to know all the environment dynamics in advance, which is an arduous task in a real world scenario. It  has also been shown that Q-Learning performs better than \textit{best fit} in all cases and it performs well considering cloud and transport network parameters, which was the main objective in the problem.


Once near-optimality of learning-based approaches is shown in this paper thanks to the mathematical tractability of the scenario, future work will be devoted to extend the proposed practical schemes towards increasingly complex 6G scenarios, e.g., through Deep Q-Networks (DQN).

\bibliographystyle{ieeetr}
\bibliography{refs}

\begin{thebibliography}{10}

\bibitem{corneo2021much}
L.~Corneo, N.~Mohan, A.~Zavodovski, W.~Wong, C.~Rohner, P.~Gunningberg, and
  J.~Kangasharju, ``(how much) can edge computing change network latency?,'' in
  {\em 2021 IFIP Networking Conference (IFIP Networking)}, pp.~1--9, IEEE,
  2021.

\bibitem{5GPPPaper}
{5G PPP Technology Board}, ``Ai and ml – enablers for beyond 5g networks,''
  2021.

\bibitem{8885745}
J.-y. Baek, G.~Kaddoum, S.~Garg, K.~Kaur, and V.~Gravel, ``Managing fog
  networks using reinforcement learning based load balancing algorithm,'' in
  {\em 2019 IEEE Wireless Communications and Networking Conference (WCNC)},
  pp.~1--7, 2019.

\bibitem{mecoff}
S.~Ranadheera, S.~Maghsudi, and E.~Hossain, ``Mobile edge computation
  offloading using game theory and reinforcement learning,'' 11 2017.

\bibitem{inproceedings}
A.~Sangpetch, O.~Sangpetch, N.~Juangmarisakul, and S.~Warodom, ``Thoth:
  Automatic resource management with machine learning for container-based cloud
  platform,'' pp.~103--111, 01 2017.

\bibitem{8647858}
L.~Yala, P.~A. Frangoudis, and A.~Ksentini, ``Latency and availability driven
  vnf placement in a mec-nfv environment,'' in {\em 2018 IEEE Global
  Communications Conference (GLOBECOM)}, pp.~1--7, 2018.

\bibitem{li20215growth}
X.~Li, A.~Garcia-Saavedra, X.~Costa-Perez, C.~J. Bernardos, C.~Guimar{\~a}es,
  K.~Antevski, J.~Mangues-Bafalluy, J.~Baranda, E.~Zeydan, D.~Corujo, {\em
  et~al.}, ``5growth: An end-to-end service platform for automated deployment
  and management of vertical services over 5g networks,'' {\em IEEE
  Communications Magazine}, vol.~59, no.~3, pp.~84--90, 2021.

\bibitem{articlewrr}
V.~Hottmar and B.~Adamec, ``Analytical model of a weighted round robin service
  system,'' {\em Journal of Electrical and Computer Engineering}, vol.~2012, 04
  2012.

\bibitem{zeydan2021}
E.~Zeydan, J.~Mangues, and Y.~Turk, ``Intelligent service orchestration in edge
  cloud networks,'' {\em IEEE Network Magazine, Interplay between Machine
  Learning and Networking System (in press)}, 2021.

\bibitem{10.5555/517430}
D.~P. Bertsekas, {\em Dynamic Programming and Optimal Control}.
\newblock Athena Scientific, 2nd~ed., 2000.

\bibitem{bakhshi2021globe}
B.~Bakhshi, J.~Mangues-Bafalluy, and J.~Baranda, ``R-learning-based admission
  control for service federation in multi-domain 5g networks,'' in {\em IEEE
  GLOBECOM}, 2021.

\bibitem{blumenfeld2001operations}
D.~Blumenfeld, {\em Operations research calculations handbook}.
\newblock CRC press, 2001.

\bibitem{sutton2018}
R.~S. Sutton and A.~G. Barto, {\em Reinforcement learning: An introduction}.
\newblock MIT press, 2018.

\bibitem{Watkins:1989}
C.~J. C.~H. Watkins, {\em Learning from Delayed Rewards}.
\newblock PhD thesis, King's College, Cambridge, UK, May 1989.

\bibitem{stochasticaprox}
H.~Robbins and S.~Monro, ``A stochastic approximation method,'' {\em The Annals
  of Mathematical Statistics}, vol.~2, no.~3, pp.~400--407, 1951.

\bibitem{qlearninghistory}
M.~Daswani, P.~Sunehag, and M.~Hutter, ``Q-learning for history-based
  reinforcement learning,'' {\em Journal of Machine Learning Research},
  vol.~29, pp.~213--228, 2013.

\end{thebibliography}

\balance
\end{document}